\documentstyle[12pt]{article}
\input{tcilatex}
\begin{document}

\author{R. Manvelyan \thanks{%
E-mail: manvel@moon.yerphi.am}, A. Melikyan, R. Mkrtchyan\thanks{%
E-mail:mrl@amsun.yerphi.am} \and {\it Theoretical Physics Department,} \and 
{\it Yerevan Physics Institute} \and {\it Alikhanyan Br. st.2, Yerevan,
375036 Armenia }}
\title{Light-Cone Formulation of D2-Brane}
\maketitle

\begin{abstract}
We apply the light-cone hamiltonian approach to D2-brane, and derive the
equivalent gauge-invariant Lagrangian. The later appears to be that of
three-dimensional Yang-Mills theory, interacting with matter fields, in the
special external induced metric, depending on matter fields. The duality
between this theory and 11d membrane is shown.
\end{abstract}

%\date{August 18 1997}

{\smallskip \pagebreak }

\section{Introduction}

The recent developments in the field of higher-dimensional extended objects
has led to the deep understanding of the non-perturbative aspects of the
superstrings and supergravity theories. That results in the unification,
through the notion of duality, of all five different superstrings theories,
and, moreover, the new, so-called M-theory \cite{MTH} notion appeared. The
M-theory is intrinsically eleven-dimensional, and contains, in the spectrum
of excitations, eleven-dimensional supermembrane theory. Also a new, in
comparison with well-known p-branes (extended objects with p space
dimensions) extended objects, so called D-branes \cite{Pol}, appeared, which
contain, in their spectrum, a vector fields (or, in the case of
eleven-dimensional 5-brane - a second rank selfdual tensor field). The main
goal of present paper is to investigate the light-cone formalism for the
bosonic part of action of D-membranes. The presence of vector field gives a
crucial difference from the known case of membranes \cite{deWitt}, and leads
to an interesting results. The supersymmetrization will be discussed
elsewhere.

The light-cone formulation of super-membrane obtained by \cite{deWitt} is
closely connected to the Matrix model representation of M-theory \cite{Banks}
. Corresponding bosonic part of area-preserving action, from which we can
obtain the Matrix model by replacing Lie brackets by commutators, looks like$%
\cite{deWitt}$:

\begin{equation}
S_{m}=\int d\tau d^{2}\sigma \left[ \frac{1}{2}(D_{0}X^{M})^{2}-\frac{1}{4}%
\left\{ X^{M},X^{N}\right\} \left\{ X^{M},X^{N}\right\} \right] ,  \label{MT}
\end{equation}
where $M=1,2,...9$ and $D_{0}=\partial _{0}+\{\omega ,...\}$ is a covariant
area-preserving derivative with gauge field $\omega (\tau ,\sigma
_{1},\sigma _{2})$ and Lie bracket

\begin{equation}
\{X,Y\}=\varepsilon ^{ij}\partial _{i}X\partial
_{j}Y\,,\,\,\,\,\,\,\,\,\,\,\,\,i,j,..=1,2.  \label{Leebra}
\end{equation}

That Lagrangian can be interpreted as a 10-dimensional Yang-Mills theory (if
we start from 11-dimensional target space for membrane) reduced to one
dimension.

The next new class of extended objects are, as mention above, the so-called
D-branes\cite{Pol}. Main feature of D-branes is that 10-dimensional
superstrings can end on them. The incorporation of D-branes in superstring
picture gives a lot of issues in the theory of solitonic states in
non-perturbative string theory and permits to reveal a different aspects of
strings/M-theory dualities.

In this article we shall investigate the light-cone formulation of the
10-dimensional D-membrane described by Dirac-Born-Infeld (DBI) Lagrangian:

\begin{equation}
L_{DBI}=-\sqrt{-G};\qquad G=\det\{g_{\mu \nu }+F_{\mu \nu }\}  \label{DBI}
\end{equation}

\[
\,g_{\mu \nu }=\partial _{\mu }X^{M}\partial _{\nu
}X^{M}\,\,\,\,,\,\,\,\,\,\,\,F=\partial _{\mu }A_{\nu }-\partial _{\nu
}A_{\mu },\, 
\]

\[
\,M=+,-,1,2,...8;\,\,\,\,\,\,\,\mu ,\nu =0,1,2, 
\]
which is a bosonic part of D2-brane. We shall construct the analog of
area-preserving membrane action (\ref{MT}) for the DBI case.

The main result of our paper is that corresponding gauged light-cone action
for D2-brane can be rewritten as a three-dimensional Maxwell theory with
matter fields, in the specific curved induced metric:

\begin{equation}
\tilde{G}_{\mu \nu }=\left( 
\begin{array}{ll}
-g+\xi ^{i}\left( \omega \right) g_{ij}\xi ^{j}\left( \omega \right) & \xi
^{k}\left( \omega \right) g_{kj} \\ 
\xi ^{k}\left( \omega \right) g_{ki} & g_{ij}
\end{array}
\right)  \label{Met}
\end{equation}
where $g_{ij}=$ $\partial _{i}X^{M}\partial _{j}X^{M},\xi ^{i}\left( \omega
\right) =\varepsilon ^{ki}\partial _{k}\omega $ and $g=\det g_{ij}$

The second result is that the duality transformation defined with this
metric tensor connects our D2-brane light-cone action to the
eleven-dimensional membrane light-cone action (connection between D2-brane
in 10d and membrane in 11d was first observed by M. J. Duff and J. X. Lu 
\cite{Duff}. Schmidhuber \cite{Sch} and Townsend \cite{THS} established this
connection in both directions). In our formulation we start from DBI action
and obtain finally some duality transformation in light-cone formulation
directly connecting that with membrane light-cone action obtained from
Nambu-Goto membrane action in 11 dimension$\left( \ref{MT}\right) .$

\section{Hamiltonian formulation}

Let's start with Hamiltonian formulation for action $\left( \ref{DBI}\right) 
$ after preliminary light-cone gauge fixing:

\begin{equation}
X^{+}(\tau ,\sigma _{i})=X^{+}(0)+\tau
,\,\,\,\,\,\,\,\,\,\,\,\,\,\,\,\,\,\,\,\,X^{\pm }=\sqrt{\frac{1}{2}}\left(
X^{10}\pm X^{0}\right)  \label{Gaug}
\end{equation}

In this gauge we have the following components of induced metric:

\begin{eqnarray}
\,G_{rs} &=&g_{rs}+F_{rs}=\partial _{r}X^{M}\partial _{s}X^{M}+\partial
_{r}A_{s}-\partial _{s}A_{r},  \nonumber \\
G_{0r} &=&g_{0r}+F_{0r}=\partial _{0}X^{M}\partial _{r}X^{M}+\partial
_{r}X^{-}+\partial _{0}A_{r}-\partial _{r}A_{0},  \nonumber \\
G_{r0} &=&g_{0r}-F_{0r}=\partial _{0}X^{M}\partial _{r}X^{M}+\partial
_{r}X^{-}+\partial _{r}A_{0}-\partial _{0}A_{r},  \label{MG} \\
G_{00} &=&g_{00}=\partial _{0}X^{M}\partial _{0}X^{M}+2\partial _{0}X^{-}, 
\nonumber \\
M &=&1,2,...8\,;\,\,\,\,\,r,s=1,2  \nonumber
\end{eqnarray}

The determinant of the induced metric and Lagrangian can be written in the
form:

\begin{eqnarray}
G &=&-\Delta \bar{G}\,;\,\,\,\,\,\,\Delta =-G_{00}+G_{0r}G^{rs}G_{s0} 
\nonumber \\
L &=&-\sqrt{\Delta \bar{G}}\,,\,\bar{G}=\det G_{rs}  \label{GL}
\end{eqnarray}

Then we can write down the canonical momenta:

\begin{eqnarray}
P^{M} &=&\frac{\partial L}{\partial \dot{X}^{M}}=\sqrt{\frac{\,\bar{G}}{%
\Delta }}\left[ \partial _{0}X^{M}-\frac{1}{2}\left( \partial
_{r}X^{M}G^{rs}G_{s0}+G_{0r}G^{rs}\partial _{s}X^{M}\right) \right] , 
\nonumber \\
P^{+} &=&\frac{\partial L}{\partial \dot{X}^{-}}=\sqrt{\frac{\,\bar{G}}{%
\Delta }},\,\,\,\,\,\,\,\,\,\,\,P_{A}^{r}=\frac{\partial L}{\partial \dot{A}%
_{r}}=\sqrt{\frac{\,\bar{G}}{\Delta }}\left[ G_{0s}G^{sr}+G^{rs}G_{s0}\right]
\label{Mom}
\end{eqnarray}

and primary Hamiltonian density and primary constraints:

\begin{eqnarray}
H &=&\frac{P^{M}P^{M}+P_{A}^{r}P_{A}^{s}g_{rs}+\bar{G}}{2P^{+}}%
+P_{A}^{r}\partial _{r}A_{0}\,,  \label{Ham} \\
P_{A}^{0} &=&0,  \label{Gm} \\
\phi _{r} &=&P^{M}\partial _{r}X^{M}+P^{+}\partial
_{r}X^{-}+P_{A}^{s}F_{rs}=0.  \label{Con}
\end{eqnarray}

Then from the requirement of conservation of primary constraints we can
obtain secondary (Gauss-low) constraint:

\begin{equation}
\chi =\partial _{r}P_{A}^{r}=0\,,  \label{GLow}
\end{equation}
corresponding to $U\left( 1\right) \,$ gauge invariance.

So we can firstly fix the gauge $A_{0}=0$ dropping the second term in $%
\left( \ref{Ham}\right) $ and correctly resolve the primary constraint $%
\left( \ref{Gm}\right) $. After that we have to add remaining first class
constraints $\left( \ref{GLow}\right) $ and$\left( \ref{Con}\right) $ to
Hamiltonian with arbitrary Lagrange multipliers $\phi _{r}$ and $\lambda $:

\begin{equation}
H=\frac{P^{M}P^{M}+P_{A}^{r}P_{A}^{s}g_{rs}+\bar{G}}{2P^{+}}+c^{r}\phi
_{r}+\,\lambda \chi  \label{Ham1}
\end{equation}

After that we can use $\tau $-dependent reparametrizations of $\sigma ^{r}:$

\[
\sigma ^{r}\longrightarrow \sigma ^{r}+\xi ^{r}\left( \tau ,\sigma
^{s}\right) 
\]
corresponding to the constraints ($\ref{Con})$, for fixing the following
gauge:

\begin{equation}
\pi _{r}=g_{0r}+\partial _{0}A_{k}g^{km}F_{rm}=0,  \label{Gaug1}
\end{equation}
where the velocities $\dot{A_{k}}$ and $\dot{X}^{M}$ have to be expressed
throw corresponding momenta and coordinates. In this and only in this gauge
after simple but tedious algebraic calculation one can prove that

\begin{eqnarray*}
c^{r} &=&0 \\
\partial _{0}P^{+} &=&0
\end{eqnarray*}
according to Hamiltonian equations of motion. It means that in analogy with
an ordinary membrane \cite{deWitt} we can put $P^{+}=1\footnote{%
Strictly speaking we have to put $P^{+}=const.\times w(\sigma _{i})$ , but
this leads only to some density factor in definition of Lie bracket \cite
{deWitt}.}$ and express $X^{-}$ coordinate through the transversal ones :

\begin{equation}
\partial _{r}X^{-}=-\left( P^{M}\partial _{r}X^{M}+P_{A}^{s}F_{rs}\right)
\label{Res}
\end{equation}

It is easy to see that after using of this expression we shall obtain the
residual constraint:

\begin{equation}
\partial _{s}\varepsilon ^{sr}\left( P^{M}\partial
_{r}X^{M}+P_{A}^{t}F_{rt}\right) =0\,  \label{Rot}
\end{equation}

Moreover, in that gauge$\left( \ref{Gaug1}\right) $ the expressions for
momenta look very simple:

\begin{eqnarray}
P^{M} &=&\partial _{0}X^{M}  \label{Mom1} \\
P_{A}^{r} &=&g^{rs}\partial _{0}A_{s}  \label{Mom2}
\end{eqnarray}
where

\begin{equation}
g^{rs}=\frac{\varepsilon ^{rt}\varepsilon ^{sp}g_{tp}}{g}  \label{met11}
\end{equation}

So, finally we obtain the following expressions for momenta, Hamiltonian and
residual constrains in light-cone gauge:

\begin{eqnarray}
P^{M} &=&\partial
_{0}X^{M},\,\,\,\,\,\,\,\,\,\,\,\,\,\,P_{A}^{r}=g^{rs}\partial _{0}A_{s}
\label{momf} \\
H &=&\frac{P^{M}P^{M}+P_{A}^{r}P_{A}^{s}g_{rs}+\bar{G}}{2},  \label{hamf} \\
\phi &=&\partial _{s}\varepsilon ^{sr}\left( P^{M}\partial
_{r}X^{M}+P_{A}^{t}F_{rt}\right) =0,  \label{conf} \\
\chi &=&\partial _{r}P_{A}^{r}=0\,  \label{gaugf}
\end{eqnarray}

\section{Gauged Lagrangian}

The main idea of this section is to find out the Lagrangian containing
fields $X^{M}$ , $A_{r}$ and two gauge field $\omega \left( \tau ,\sigma
_{i}\right) $ and $Q(\tau ,\sigma _{i})$ with the following properties:

1.The expressions $\left( \ref{momf}\right) $ and $\left( \ref{hamf}\right) $
have to be derived as standard expressions of the canonical momenta and
Hamiltonian for that Lagrangian in the gauge

\begin{eqnarray}
\omega \left( \tau ,\sigma _{i}\right) &=&0  \nonumber \\
Q(\tau ,\sigma _{i}) &=&0  \label{gaug2}
\end{eqnarray}

2.The equation of motion for gauge fields $\omega \left( \tau ,\sigma
_{i}\right) $ and $Q(\tau ,\sigma _{i})$ have to coincide (in the gauge $%
\left( \ref{gaug2}\right) $) with corresponding constraints $\left( \ref
{conf}\right) $ and$\left( \ref{gaugf}\right) $.

3.This Lagrangian has to be gauge invariant with following gauge groups:

a)group of area-preserving diffeomorphisms corresponding to the constraint $%
\left( \ref{conf}\right) $:

\begin{equation}
\phi =\left\{ \partial _{0}X^{M},X^{N}\right\} \,+\varepsilon ^{sr}\partial
_{s}\left( \partial _{0}A_{k}g^{kt}F_{rt}\right) =0  \label{constl}
\end{equation}

b)group of $U\left( 1\right) $ gauge transformations connected to constraint 
$\left( \ref{gaugf}\right) $:

\begin{equation}
\chi =\partial _{r}\left( g^{rs}\partial _{0 }A_{s}\right) =0  \label{gaugl}
\end{equation}

The desired Lagrangian has the following form:

\begin{equation}
L=\frac{\left( D_{0}X^{M}\right) ^{2}}{2}+\frac{1}{2}g^{rs}\left(
D_{0}A_{r}-\partial _{r}Q\right) \left( D_{0}A_{s}-\partial _{s}Q\right) -%
\frac{\bar{G}}{2}\,,  \label{Lag1}
\end{equation}
where $\bar{G}=g+F_{12}^{2}$ and

\begin{eqnarray}
&&D_{0}X^{M}=\partial _{0}X^{M}-\varepsilon ^{ij}\partial _{i}\omega
\partial _{j}X^{M}=\partial _{0}X^{M}-\left\{ \omega ,X^{M}\right\}
=\partial _{0}X^{M}-\pounds _{\xi \left( \omega \right) }X^{M}  \nonumber \\
&&D_{0}A_{r}=D_{0}A_{r}=\partial _{0}A_{r}-\varepsilon ^{ij}\partial
_{r}\partial _{i}\omega A_{j}-\varepsilon ^{ij}\partial _{i}\omega \partial
_{j}A_{r}=\partial _{0}A_{r}-\pounds _{\xi \left( \omega \right) }A_{r}.
\label{CovD}
\end{eqnarray}
Here $\pounds _{\xi \left( \omega \right) }$ is the Lie derivative in
direction of divergenceless vector field $\xi ^{i}\left( \omega \right)
=\varepsilon ^{ki}\partial _{k}\omega .$

Lagrangian $\left( \ref{Lag1}\right) $ satisfies all three conditions, the
gauge transformations for given fields are following:

\begin{eqnarray}
\delta _{\varepsilon }X^{M} &=&\left\{ \varepsilon ,X^{M}\right\} ,\delta
_{\varepsilon }A_{r}=\pounds _{\xi \left( \omega \right) }A_{r},  \nonumber
\\
\,\delta _{\varepsilon }Q &=&\left\{ \varepsilon \,,Q\right\} ,\delta
_{\varepsilon }\omega =\partial _{0}\varepsilon +\left\{ \varepsilon
\,,\omega \right\} ,  \label{dtr1} \\
\delta _{\alpha }X^{M} &=&0,\delta _{\alpha }A_{r}=\partial _{r}\alpha
,\delta _{\alpha }Q=\partial _{0}\alpha +\left\{ \alpha \,,\omega \right\}
,\delta _{\alpha }\omega =0.  \label{gtr1}
\end{eqnarray}

It is easy to see that $U\left( 1\right) $ gauge transformations of $Q$ do
not commute with area-preserving ones.

This can be improved by redefinition of field $Q$:

\begin{equation}
A_{0}=Q+\varepsilon ^{ij}\partial _{i}\omega A_{j}  \label{azero}
\end{equation}

Here we introduce new $A_{0}$ component, which, differently from $Q,$
transforms (under area-preserving diffeomorphisms) not as a scalar but as a
zero component of three-dimensional vector field :

\begin{eqnarray}
\delta _{\alpha }A_{0} &=&\partial _{0}\alpha  \nonumber \\
\delta _{\varepsilon }A_{0} &=&\left\{ \varepsilon \,,A_{0}\right\}
+\partial _{0}\xi ^{i}\left( \varepsilon \right) A_{i}  \label{gtr2}
\end{eqnarray}
After that the Lagrangian $\left( \ref{Lag1}\right) $ can be rewritten in
the following form:

\begin{eqnarray}
L &=&\frac{\left( D_{0}X^{M}\right) ^{2}}{2}-\frac{1}{4}\left\{
X^{M},X^{N}\right\} \left\{ X^{M},X^{N}\right\} +\frac{1}{2}%
g^{ij}F_{0i}F_{0j}  \label{Lag2} \\
&+&\frac{1}{2}g^{ij}\xi ^{m}\left( \omega \right) \xi ^{n}\left( \omega
\right) F_{im}F_{jn}-\frac{1}{2}F_{12}^{2}+\frac{1}{2}g^{ij}F_{0i}F_{jn}\xi
^{n}\left( \omega \right) +g^{ij}\xi ^{m}\left( \omega \right) F_{im}F_{0j} 
\nonumber
\end{eqnarray}
here $F_{0r}=\partial _{0}A_{r}-\partial _{r}A_{0}$ .

Therefore, after introduction of three-dimensional metric $\tilde{G}_{\mu
\nu }$ $\left( \ref{Met}\right) $ with following properties:

\begin{eqnarray}
\tilde{G}_{\mu \nu } &=&\left( 
\begin{array}{ll}
-g+\xi ^{i}\left( \omega \right) g_{ij}\xi ^{j}\left( \omega \right) & \xi
^{k}\left( \omega \right) g_{kj} \\ 
\xi ^{k}\left( \omega \right) g_{ki} & g_{ij}
\end{array}
\right) ,  \nonumber \\
\tilde{G}^{\mu \nu } &=&\left( 
\begin{array}{ll}
-1/g & \xi ^{j}\left( \omega \right) /g \\ 
\xi ^{i}\left( \omega \right) /g & g^{ij}-\xi ^{i}\left( \omega \right) \xi
^{j}\left( \omega \right) /g
\end{array}
\right) ,  \label{Met2} \\
\,g_{ij} &=&\partial _{i}X^{M}\partial _{j}X^{M},\,g=\det g_{ij}\,,\,\xi
^{i}\left( \omega \right) =\varepsilon ^{ki}\partial _{k}\omega  \nonumber \\
\det \tilde{G}_{\mu \nu } &=&\tilde{G},\,\,\sqrt{-\tilde{G}}%
=g,\,\,\,\,\,\,\,\,\sqrt{-\tilde{G}}G^{00}=-1  \nonumber
\end{eqnarray}

and using $\left( \ref{met11}\right) $we can obtain from $\left( \ref{Lag2}%
\right) $ the final expression for our effective

light-cone action:

\begin{eqnarray}
L &=&-\frac{1}{2}\sqrt{-\tilde{G}}\tilde{G}^{\mu \nu }\partial _{\mu
}X^{M}\partial _{\nu }X^{M}+\frac{1}{2}\sqrt{-\tilde{G}}  \nonumber \\
&&-\frac{1}{4}\sqrt{-\tilde{G}}\tilde{G}^{\mu \nu }\tilde{G}^{\sigma \lambda
}F_{\mu \sigma }F_{\nu \lambda },  \label{Lagf} \\
\partial _{\mu } &=&\left( \partial _{0},\partial _{i}\right) ,\,\,F_{\mu
\sigma }=\left( F_{0r},F_{ij}=F_{12}\varepsilon _{ij}\right) ,  \nonumber
\end{eqnarray}

Here we used relation:

\[
\left\{ X^{M},X^{N}\right\} \left\{ X^{M},X^{N}\right\} =\sqrt{-\tilde{G}}%
\left( \tilde{G}^{ij}+\xi ^{i}\left( \omega \right) \xi ^{j}\left( \omega
\right) /g\right) \partial _{i}X^{M}\partial _{j}X^{M} 
\]

So, we proved that effective action for light-cone 10d D2-brane can be
expressed in the form of usual three-dimensional abelian gauge field coupled
to eight scalar matter field $X^{M}$ in the induced metric $\left( \ref{Met2}%
\right) $defined by the same matter fields - target space coordinates $%
X^{M}. $

\section{Duality transformation}

Let us introduce the new coordinate $X^{9}$ by standard Abelian duality
transformation with our metric $\left( \ref{Met2}\right) $.

For that let us add to $\left( \ref{Lagf}\right) $ metric independent
topological term :

\begin{equation}
\int L\left( X,\omega ,F\right) d\tau d^{2}\sigma +\frac{1}{2}\int X^{9}dF
\label{GenLag}
\end{equation}

Here $F$ is independent second rank antisymmetric tensor field.

Integration over $X^{9}$ leads to $\left( \ref{Lagf}\right) .$ But
integration over $F$ gives the following equation of motion:

\begin{equation}
\partial _{\mu }X^{9}=\sqrt{-\tilde{G}}\varepsilon _{\mu \nu \lambda }\tilde{%
G}^{\nu \rho }\tilde{G}^{\lambda \sigma }F_{\rho \sigma }  \label{dual1}
\end{equation}

or in components:

\begin{eqnarray}
\partial _{0}X^{9} &=&F_{12}  \nonumber \\
\partial _{i}X^{9} &=&\varepsilon _{ij}\left( g^{jk}-\xi ^{j}\left( \omega
\right) \xi ^{k}\left( \omega \right) /g\right) F_{0k},  \label{dual2}
\end{eqnarray}

We see that the substitution of $\left( \ref{dual2}\right) $ in $\left( \ref
{GenLag}\right) $ leads to:

\[
L=\frac{1}{2}(D_{0}X^{M})^{2}-\frac{1}{4}\left\{ X^{M},X^{N}\right\} \left\{
X^{M},X^{N}\right\} 
\]

where $M,N,.=1,2,...9.$

This is the light-cone effective Lagrangian for eleven-dimensional membrane.
As mentioned above, connection between 10d D2-brane and 11d membrane was
established and exploited in \cite{Duff}, \cite{THS}, \cite{Sch}.

\section{Conclusion}

In the present paper the light-cone formalism is developed for
10-dimensional D2-brane, and it was shown, that all corresponding equations
of motion and constraints can be derived from the Lagrangian of usual 3d
Maxwell theory, interacting with matter fields, in a curved space-time with
a special induced metric. This theory is invariant with respect to usual
Abelian gauge transformations of gauge fields, and with respect to
area-preserving diffeomorphisms. So, we have shown, that complicated
non-linear DBI Lagrangian can be substituted, at least at classical level,
in light-cone gauge, with quadratic one over gauge fields, although the
dependence on a matter fields (coordinates of membrane) remains highly
non-linear. There is no any direct connection to a small (gauge) field
expansion of initial DBI Lagrangian, although they seem to be similar (but
of course there is a lot of differences). The exact integration over gauge
fields can be carried out now, at least formally, the corresponding
determinant has to be considered as an effective action for D-brane, and can
be expanded by the Riemann tensor (and it's derivatives) of the metric $%
\left( \ref{Met2}\right) $. The properties of that tensor, with given
special metric, may be very peculiar. Another way of thinking (which was an
initial motivation of the present study) is possible connection with Matrix
models. Unfortunately, there is no any evident way of interpreting fields in
Lagrangian $\left( \ref{Lagf}\right) $ as a matrixes, with corresponding
immersion of gauge groups. Nevertheless, there are some indications that
literally same Lagrangian can be derived for D3-brane. That problem,
together with supersymmetrization of these results will be considered in
separate paper\cite{MMM}

{\bf Acknowledgments}

This work was supported in part by the U.S. Civilian Research and
Development Foundation under Award \# 96-RP1-253 and by INTAS grants \#
96-538 and \# 93-1038 (ext) .

\end{document}